# Скелетные структуры океана, гипотезы и интерпретация явления.

В.А.Ранцев-Картинов

ИЯС РНЦ "Курчатовский Институт", Москва 123182, Россия
rank@nfi.kiae.ru

В работе выдвинута гипотеза о формировании скелетных структур океана (ССО) из фрагментов каркасных структур (КС) облаков при попадании их на поверхность океана в результате атмосферных процессов. Эти КС самопроизвольно собираются из структурообразующей пыли (СОП) под действием атмосферного электричества. Предполагается, что СОП содержит углеродные нанотрубки или подобные им частицы и поставляется в атмосферу вулканами при их извержении. Высказана гипотеза о динамике преобразования КС в ССО во время штормов. Дана интерпретация наблюдаемому явлению на основе действия сил поверхностного натяжения. Особое внимание уделено блокам ССО в виде вертикально плавающих цилиндров. На их примере оценена прочность и плавучесть отдельных блоков ССО. Высказано предположение о динамике ССО в штилевой период после шторма, определяющей локальную отражательную способность поверхности океана, наблюдаемую из космоса.

**1. Введение.** Анализ фотоизображений поверхности океана, полученных с различных высот и при различных уровнях волнения его поверхности, привел автора к наблюдению ССО [1]. КС, подобные ССО, с такой же топологией и с теми же свойствами (долгоживучестью, самоподобием и стремлением к самосборке) ранее уже наблюдались в целом ряде явлений в лаборатории, атмосфере и космосе [2-4]. В публикациях [2c,d,f] были подробно описаны и даны объяснения некоторым свойствам таких КС и предложены гипотезы, основа которых состояла в том, что *только микропылевая компонента с её квантовыми связями может быть ответственна за наблюдаемые в плазме КС* [2c]. Итоговые работы по КС, [2e,3a], дали выводы *о роли нанопыли в формировании КС в широком диапазоне масштабов*, охватывающих почти 30 порядков величины.

**2. Гипотезы.** В [1] автором были выдвинуты гипотезы по формированию ССО. Основные их положения можно представить следующим образом: *a) - в основе ССО* (как и описанных ранее КС [2-4]) *лежит структурообразующая пыль, базовыми элементами которой могут являться углеродные нанотрубки (УНТ) или подобные им частицы из других элементов, поставляемые в атмосферу вулканами; b) - под воздействием электрического поля Земли и атмосферного электричества такая пыль может выстраивать КС облаков, фрагменты которых в результате различных атмосферных явлений попадают на поверхность океана; c) - КС имеют очень активную поверхность и не нулевую плавучесть, поэтому, адсорбируя на себе растворенный в воде воздух или частично заполняясь пеной, всплывают к поверхности; d) - создавая на своей поверхности разделение трех* *фаз вещества (твердой, жидкой и газообразной), КС обеспечивают действие сил поверхностного натяжения (благодаря неполному смачиванию) и сцепление между собой отдельных блоков в единую ССО, даже под водой; **e)** - прочность ССО и отдельных ее блоков определяется тем, что они плотно упаковываются фрагментами КС попавших на поверхность океана и блоками предыдущих поколений; **f)** - скелетные свойства ССО проявляется в том, что довольно жесткие отдельные блоки ее гибко и мягко (с возможностью растяжения, благодаря силам поверхностного натяжения(СПН)) сочленяются друг с другом, как в скелете.*

**3. Наблюдения ССО.** В [1] дан краткий перечень типов наблюдаемых в ССО блоков с их кратким описанием и фрагментами изображений. Одним из выводов этой работы был *факт увеличения основных размеров наблюдаемых блоков ССО по мере увеличения волн*. В данной работе будут проведены оценки прочности сцепления и плавучести блоков ССО в виде вертикально плавающих цилиндров (ВПЦ), которые представляют собой коаксиально-трубчатые (КТ) блоки плотно набитые во время штормов фрагментами КС попавших на поверхность океана и блоками предыдущих поколений. В таком случае они могут оказаться сплошь/частично заполненными тонкими капиллярами в виде УНТ различных поколений или подобными частицами из других элементов. Такие капилляры заполняют блоки любых типов ССО. Конструкция отдельных блоков ССО в рамках своей топологии могут быть различной ажурности. При этом ажурность блоков может не вести к потере их прочности. Такой, например, КТ блок с «тележными колесами» на торцах и параллельными спицами вдоль образующей цилиндра представляет собой КТ блок типа беличьего колеса. Имея максимальную легкость конструкции, такой блок очень прочен и устойчив в сохранении своей формы. При этом все силовые элементы его могут быть собраны из вышеупомянутых капилляров. КТ блок может иметь гладкую боковую поверхность или заполняться трубами меньшего диаметра, которые, в свою очередь, составлены из капилляров. Капилляры могут в разной степени заполняться водой с воздухом (пеной), обеспечивая тем самым некоторую плавучесть всей ССО в целом. Для демонстрации ВПЦ приведем пример из [1], который является фрагментом фото полученного летающей лабораторией во время шторма Белле с высоты 500 футов [5] и представленным здесь на рис.1. При анализе исходного фрагмента с целью выявления четкости наблюдаемой структуры ВПЦ применен метод многоуровневого динамического контрастирования, ранее разработанный и описанный автором [2a,b,3a].

**4. Интерпретация наблюдаемого явления.** Указанная выше гипотеза **d** отмечает - наличие среды в ином фазовом состоянии может внести существенные коррективы в уже рассмотренные ранее процессы формирования КС [2e, 3a,]. Попробуем оценить: "Достаточно ли СПН и капиллярных явлений для объяснения наличия наблюдаемых в волнах ВПЦ, которые заполнены водой с пеной и возвышаются над средним уровнем вплоть до метра?" Положим, что во время шторма есть силы, способные вытолкнуть конструкцию из тонких капилляров над средним уровнем воды на достаточно большую высоту, а стенки капилляров смачиваемы. Тогда из теории СПН можно оценить высоту подъема жидкости в них. Если конструкция ВПЦ заполнена капиллярами расположенными параллельно его образующей и вес столба жидкости уравновешен СПН, то нет и силы действующей на разрыв структуры блока, а сам вес уравновешиваться силами ответственными за плавучесть блока. В промежутках между заполняющими блок капиллярами образуются капилляры, характерный размер которых меньше основных. Поэтому в таком идеальном блоке вес всей жидкости ВПЦ поднятой над средним уровнем уравновешен капиллярными силами. Такой блок будет сохранять свою форму, и его торцевая поверхность будет омываться водой при волнении поверхности океана как сплошная/ажурная поверхность, в зависимости от конкретной его конструкции. Итак, если объем ВПЦ при волнении частично заполнился пеной, то он всплывает на некоторую высоту над средним уровнем. Капиллярные силы держат воду в своих каналах. Блок остается наплаву, сохраняя свою геометрическую форму с вертикальной боковой поверхностью, пока пена не выйдет из капилляров. Заполнение пеной ВПЦ обусловлено опрокидыванием волны на его торец во время шторма, что имеет место даже в открытом океане. В вертикальном направлении жидкость во всплывших капиллярах натянута ее весом. В радиальном направлении структура ВПЦ имеет прочностные характеристики цилиндрического жгута, составленного из тонких параллельных капилляров связанных между собой СПН. Легко показать, что соответствующий модуль Юнга $E$ для ВПЦ можно оценить, как $E = 4\sigma/d$ г/см², где $\sigma$ - коэффициент СПН для воды и $d$ - диаметр капилляров наполняющих тело ВПЦ. Т.е., чем тоньше капилляры, тем прочнее конструкция ВПЦ. Легко показать, что в случае отклонения оси ВПЦ от вертикали на угол $\alpha$, заполненная жидкостью длина капилляра $l$ пропорциональна: $1/\cos\alpha$. А тогда, при наклонах такого блока в волнах, жидкость будет выливаться из капилляров блока, а сам блок подтапливаться и терять устойчивость с частичным разрушением его наклонённой боковой поверхности, что иногда и наблюдается.

    Оценим теперь, какой величиной сверху ограничивается *диаметр базовых капилляров* заполняющих объем таких ВПЦ и способных поднять жидкость на высоту, $h$. Эта оценка следует из неравенства: $d \leq 4\sigma/gh\rho$, где $g$ - ускорение свободного падения, $\rho$ - плотность воды. Подстановка сюда справочных данных даёт значение предельного диаметра капилляра, способного транспортировать жидкость в гравитационном поле на заданную высоту. Если $h \sim 1$ м, тогда ограничение на диаметр капилляра определяется величиной $d \leq 3 \cdot 10^{-3}$ см. При достаточно большой плавучести (способной держать столб воды, поднятый над средним уровнем за счёт капиллярных сил на высоту $h$) подобная капиллярная конструкция будет сохранять свою цилиндрическую форму и плавать вертикально, как поплавок. Оценим теперь плавучесть ВПЦ и способность их нести нагрузку в виде столба жидкости диаметром равным внешнему диаметру плавающей структуры. Для оценки необходимо сначала задаться некоторой моделью построения самоподобных трубчатых структур. Базовым блоком для нашего построения выберем одно/двухслойную УНТ длинной ~ 100 Å. Такие базовые УНТ с большой вероятностью получаются при любом энергетическом воздействии на кристаллический углерод. Кристаллические плоскости углерода

представляет собой мозаичные плоские ковры из лоскутов (с характерным размером ~ 100 Å и плотно выложенными шестигранниками со стороной ~ 1,42 Å, а в узлах их расположены атомы углерода) с ослабленными связями между ними. УНТ первого поколения – это свернутый в свиток или трубку такой лоскут, содержащий ~ 8 10$^3$ атомов углерода и массой $m_1$ ~ 1,7 10$^{-19}$ г. Далее выложим подобный же лоскут нанотрубками 1-го поколения и свернем его в трубку второго поколения, и т.д. Масса, длина и диаметр трубок n-го поколения определяются в таком случае соответственно: $m_n \sim 1,7 \cdot 10^{-19} \cdot 10^{4(n-1)}$ г, $L_n \sim (75)^n \cdot 1,42 \cdot 10^{-8}$ см, $D_n \sim 3 \cdot 10^{-7} \cdot (75)^{n-1}$. Торцы УНТ до 3-го поколения могут закупориваться плоскими чешуйками фитопланктона, имеющих характерный размер ~ 5 10$^{-4}$ см и обеспечивающих тем самым плавучесть ССО. Между этим размером и диаметром капилляра, способного поднять водяной столб на высоту ~ 1 м, имеется зазор порядка величины. Кроме того, УНТ, способны интенсивно адсорбировать из воды растворённый в ней газ и (по мере формирования и собирания его пузырьков на своей поверхности) создавать дополнительную плавучесть блокам ССО. Оказывается, что только УНТ 3-го поколения способны вытягивать водяной столб на метровую высоту. Действительно, $D_3 \sim 1,7 \cdot 10^{-3}$ см, что соответствует этому условию, но они не закупориваются частицами фитопланктона, поскольку они в несколько раз меньше полученных значений диаметров. Удельный вес выбранного нами блока КС, диаметр которого ~ 8 м, определяется номером его поколения. Он легко определяется и равен n ~ 6. Отсюда легко рассчитать плотность каркаса такой трубы ~ 1,5 10$^{-5}$ $\rho$, поэтому ей в оценках можно пренебрегать. При сильном шторме прочности исходной КС недостаточно, чтобы выстоять под ударами разбушевавшейся стихии. Как следует из визуального анализа изображений поверхности во время штормов, этот каркас дробиться и выстраивается более прочная структура, которая представляет собой трубы, соответствующие найденному поколению, но плотно заполненные трубками намного меньшего диаметра (другой предельный случай). Причём, как видно из визуального анализа, при таком заполнении внутри могут оставаться незаполненными капиллярами трубы предыдущих поколений. Теперь попробуем описать модификацию первоначальной КС в морской воде и превращения ее в ССО. Из выше сказанного, нас могут удовлетворить только трубки 3-го поколения, следовательно, в построении ВПЦ могут участвовать именно такие трубки. Если интенсивное волнение океана разрушает часть попавших на его поверхность КС, то его поверхность оказывается выстланной сплошным ковром из этих трубок. Поскольку при сильном ветре волны имеют преимущественную ориентацию, то и трубки выстраиваются вдоль гребней волн. Они удлиняются, устанавливая связи между собой за счет СПН, выстилая поверхность сплошным ковром. Анализ изображений показывает, что длина подобных нитей (и даже значительно большего диаметра) могут достигать порой сотни метров. Наращивание волнения моря приводит к сворачиванию этого ковра в цилиндрические рулоны (диаметр которых отвечает размеру максимальной высоты наблюдаемых в данный момент волн) из плотно упакованных трубок меньшего диаметра, одновременно набивая этим капиллярным "мусором" уцелевшие блоки первоначальных КС. Статистика визуального исследования топологии таких наблюдаемых рулонов, может дать в нулевом приближении некоторое правило их построения – цилиндр данного поколения представляет собой связку из цилиндра предыдущего поколения и, чаще всего, 6-ти таких же цилиндров вокруг него заключённых в одну цилиндрическую оболочку. Причём, оставшиеся при таком построении свободные промежутки могут заполняться или не заполняться цилиндрами более ранних поколений. Суммарный процент заполнения может оказаться близким к 100 %. Указанное правило выстраивания поколений в таком процессе является вполне естественным, ибо процесс самосборки и укрупнения цилиндров идёт параллельно с процессом наращивания высоты волн и их энергетической мощи. При интенсивном волнении поверхности и опрокидывания волн идёт взбивание морской пены, которая играет дополнительную роль в увеличении плавучести формируемых при этом каркасных блоков, т.к. она способствует уменьшению среднего удельного веса воды, заполняющей тела этих блоков составляющих ССО. Оценим теперь плотность углерода в каркасном блоке максимального наблюдаемого диаметра ~ 8 м. при условии, что он полностью упакован УНТ 3-го поколения. Оценка даёт величину ~ $3 \cdot 10^{-4} \rho$, что в ~ 20 раз превышает плотность конденсированной компоненты в воде первоначально идеальной модели КС. Оказывается, что плотность пылевой компоненты в ССО вблизи поверхности воды при усилении волнения возрастает, но она вносит ничтожно малый вклад в усреднённое значение плотности морской воды, поскольку содержание солей и других примесей в ней почти в $10^5$ раз превышает долю обсуждаемой нами пылевой компоненты. Поэтому выявить её присутствие в воде простым выпариванием и разделением весьма проблематично. О наличие её в водах океана можно судить только путём очень тонкого физического/химического анализа или анализа особых свойств и явлений, проявляющихся при различных уровнях волнения океана, равно как и в аномальных атмосферных явлениях, связанных с морем.

    Попробуем теперь оценить плавучесть максимального диаметра наблюдаемого и уже описанного выше цилиндра наполненного водой и возвышающегося над средним уровнем воды некоторое, но конечное время. Положим, мы уже убедились в наличие ВПЦ и приняли его к рассмотрению, как факт. Из наблюдений следует, что для подобных, но горизонтально плавающих цилиндров (ГПЦ) и наблюдающихся при очень сильных штормах длина может достигать $L \sim (50 - 100)$ м. Этого оказывается уже достаточно для оценки усреднённой по объёму ВПЦ плотности наполняющей его воды, чтобы он всплыл. Для ВПЦ, заполненного жидкостью со средней плотностью $\rho_a$ и всплывшего над средним уровнем на высоту $h \sim 1$ м, можно записать соотношение: $\rho/\rho_a = (L-h)/L$. Подставляя сюда численные значения $L$ и $h$, получаем значение $\rho_a = (0{,}98 - 0{,}99)\rho$. Т.е., суммарный объём воздуха в воде, заполняющей плавающий цилиндр, составляет всего (1-2) %. По мере выхода этих пузырьков воздуха в атмосферу и выхода вспененной воды из капилляров ВПЦ будет подтапливаться, и опускаться к поверхности. Аномальная плавучесть ВПЦ существует, благодаря разности времени выхода воздушных пузырьков из открытой воды и капилляров и пока доля воздуха

в воде цилиндра не сравняется с таковой в окружающей воде. Повышенная же концентрация таких пузырьков появляется при опрокидывании гребня набегающей волны на вертикально ориентированный и подтопленный цилиндр или при поступлении вспененной воды в него снизу. Это имеет место при проникновении мощных струй вспененной воды на большую глубину и заполнении ей ВПЦ у поверхности при ее всплывании. После чего сам ВПЦ всплывает на некоторое время. Итак, *УНТ компонента обеспечивает действие СПН даже под водой, тем самым, создавая силовую гарантию сохранности созданных блоков ССО и их плавучесть.* Суммарная сила сцепления ВПЦ над поверхностью воды может оказаться весьма значительной (из-за сильной развитости общей взаимодействующей боковой поверхности блока), поскольку он плотно заполнен взаимодействующими тонкими трубками. Предельное значение погонной плотности энергии такого взаимодействия пропорционально отношению $(D/d)^2$, где $D$ - диаметр блока и $d$ - диаметр трубок его заполняющих.

**5. Заключение.** Следует отметить, что физическое ощущение наличия в воде ССО такое же, как если бы Вы опускали руку в воду, поскольку кроме сил поверхностного натяжения здесь других не проявляется. Что же происходит с ССО после прекращения шторма? Часть из них уйдёт на некоторую глубину, и ничем не будет проявлять себя до следующего шторма, способного извлечь их на поверхность. Другая же, особенно ГПЦ, будет слегка возвышаться над поверхностью. Выступающие из воды их части под действием солнца будут высыхать, терять связующие их жидкие пленки и рассыпаться на составляющие фрагменты, которые, в свою очередь, будут подтапливаться или плавать на поверхности, в зависимости от их индивидуальной плавучести. Такие плавающие фрагменты рассмотренной ССО имеют тенденцию к выстраиванию структуры, соответствующей уровню волнения моря в данный момент и в данном районе. Очевидно, что для каждого уровня волнения и других, как погодных, так и физических условий, время формирования более или менее стабильной и однородной сети структуры будет индивидуальным и конечным. Поэтому понятно, почему из космоса в штилевую погоду хорошо просматриваются следы маршрутов кораблей порой до полутысячи километров, т.е. в течение суток, ибо отражательная способность поверхности океана, зависит от наличия, вида структур и их характерных масштабов. Действительно, винты кораблей размалывают и разрушают однородность установившейся структуры, а время её восстановления требует времени и идентичности истории установления её. Т.е., *индивидуальность структуры в данном районе определяется предысторией её становления.*

Литература

[1] V.A.Rantsev-Kartinov, http://www.arxiv.org/ftp/physics/papers/0401/0401139.pdf
[2] A.B.Kukushkin, V.A.Rantsev-Kartinov, a) Laser and Particle Beams, **16**, 445,(1998); b) Rev.Sci.Instrum., **70**, 1387,( 1999); c) Proc. 17-th IAEA Fusion Energy Conference, Yokohama, Japan, **3,** 1131, (1998); d) Proc. 26-th EPS PPCF, Maastricht, Netherlands, 873, (1999); e) Phys.Lett. A, **306**, 175; f) Current Trends in International Fusion Research: Review and Assessment (Proc. 3[rd] Symposium, Washington D.C., March 1999), Ed. E. Panarella, NRC Research Press, Ottawa, Canada, 121, (2002); g) "Advances in Plasma Phys. Research", (Ed. F. Gerard, Nova Science Publishers, New York), **2**, 1, (2002); h) Preprint of Kurchatov Institute, IAE 6111/6, Moscow, October (1998), Sec. 6.
[3] Кукушкин А.Б., Ранцев-Картинов В.А., a) Наука в России, 2004, **1**, 42, (2004); b) Микросистемная техника, **3**, 22 (2002); c) Мат. IV Российского семинара "Современные средства диагностики плазмы и их применение для контроля веществ и окружающей среды", Москва, МИФИ, 151, (2003).
[4] B.N.Kolbasov, A.B.Kukushkin, V.A.Rantsev-Kartinov, et.set., *Phys. Lett. A:* a) **269**, 363, (2000); b) **291,** 447, (2001); c) Plasma Devices and Operations, **8** , 257, (2001).
[5] http://www.photolib.noaa.gov/flight/images/big/fly00164.jpg

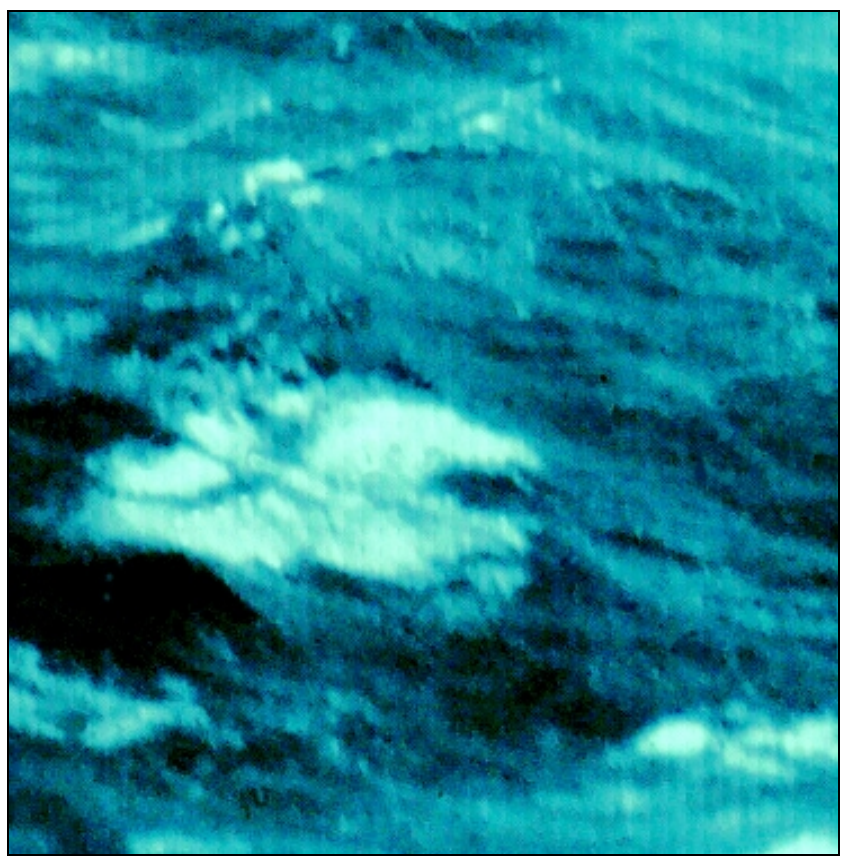

Рис.1. Фрагмент фото [5], ширина рисунка соответствует ~ 15 м. На переднем плане рисунка видна почти половина диаметра ВПЦ, ось которого наклонена вправо от вертикали ~ 30°. Диаметр ВПЦ~ 10 м, диаметр центральной осевой трубки ~ 2 м, диаметры темных окружностей на торце ВПЦ ~1 м. ВПЦ всплыл над средним уровнем на ~ 1 м.